# The End of the Foundation Model Era:

# Open-Weight Models, Sovereign AI, and Inference as Infrastructure

*A Structural Analysis of AI's 2026 Transition Across Economic, Technical, and National Security Dimensions*

---

Jared James Grogan

*Universitas AI*

March 9, 2026

*Version 1.0 — Events covered through March 9, 2026*

*Preprint. All citations carry confirmed URLs unless otherwise noted in the bibliography entry. The structural arguments do not depend on any single citation.*

---

## Abstract

The foundation model era — roughly 2020 to 2025 — is over. The forces that defined it have inverted. Open source models have reached frontier performance while inference costs approach zero, exposing what was always structurally true: pre-training large language models at scale is not a durable competitive moat. The US government's formal designation of Anthropic as a supply chain risk in February 2026 accelerated a transition already underway — but did not cause it. The paper argues that the AI industry is restructuring simultaneously along four axes: economic, as the circular financing structure that inflated foundation model valuations collapses; technical, as the pre-training scaling paradigm gives way to post-training optimization, test-time compute, and agentic composition; commercial, as application-layer integrators displace the foundation model companies whose commodity they now consume; and political, as the government asserts its historic role as gatekeeper of strategic technology. These are not separate disruptions. They are one structural shift, arriving together. The most consequential and least-discussed dimension is the permanent divergence between commercial AI and a classified national security AI track — built on different data, governed by different rules, and developing capabilities the public ecosystem cannot see, measure, or govern. Like every dual-use technology that has altered the calculus of state power, AI is being brought under government authority not by design but by the structural logic of what it is. The paper further argues that open-weight models are the counterintuitive instrument of sovereign control: a government that holds the weights commands the capability on its own terms, without dependence on vendor policy, financial continuity, or personnel clearance. The apparent openness of distributed model weights is, from a deploying government's sovereignty standpoint, the most governable architecture — because what cannot be withdrawn by a vendor's API policy cannot be taken away. The paper draws on public financial disclosures, primary government and corporate sources, and academic literature in AI scaling, innovation economics, and national security technology policy.

## I. The 2026 Inflection Point

Between late 2025 and early 2026, the AI industry crossed a threshold that had been approaching for years. The US government's formal designation of Anthropic as a supply chain risk — under both DoD supply chain authorities (10 U.S.C. § 3252) and federal ICT supply chain security frameworks (41 U.S.C. § 1323) — and its exclusion from government contracts were the most visible markers [1][2]. The Department of War[1] and the White House made clear that AI for national security applications would not remain in the discretionary control of private companies operating under their own usage policies and their own financial constraints.

But the government action did not initiate the structural change. It confirmed it. The economics of pre-training had already deteriorated. Open source models had already reached performance parity with commercial offerings for most tasks. Investment was already moving from foundation model companies toward the application layer. The Anthropic designation arrived into a transition already underway — and it matters analytically to keep these forces separate, because each is sufficient on its own. Together, they are decisive.

The foundation model era is over. What it means to understand AI — as industry, as technology, as strategic asset — must be recalibrated accordingly. Anthropic's legal challenge to its designation, filed March 9, 2026, is examined in Section III; the structural analysis is independent of its outcome.

## II. The Commoditization of Foundation Models

### The Technical Ceiling

The pre-training paradigm rested on a straightforward empirical observation: as models grew larger and training data increased, performance improved in roughly predictable ways. The scaling laws documented by Kaplan et al. [3] and refined by Hoffmann et al. [4] gave this a quantitative foundation. Each model generation — GPT-2 to GPT-3, GPT-3 to GPT-4 — appeared to validate the extrapolation. The assumption was that the trajectory would continue indefinitely: that frontier performance was simply a function of compute, data, and capital.

It is not. The benchmark record makes the pattern concrete. The jump from GPT-2 to GPT-3 was qualitative: GPT-3 introduced emergent few-shot learning — the ability to perform novel tasks without task-specific fine-tuning — and produced benchmark improvements of twenty to sixty percentage points across language tasks [5]. GPT-3's 175 billion parameters, approximately 117 times GPT-2's 1.5 billion, yielded a step change in kind, not merely degree. The jump from GPT-3 to GPT-4 was still significant but narrower in character: benchmark improvements of fifteen to twenty percentage points on major evaluations — MMLU improved from roughly 70% to 86.4%, HumanEval coding benchmark from 48% to 67% — at a training cost estimated well over an order of magnitude higher [6][7]. The capability curve was flattening while the cost curve was steepening: the same dollar of additional compute was buying a progressively smaller increment of measurable capability advantage over the prior generation.

The post-GPT-4 generation extended the pattern and made it impossible to ignore. As the leading frontier labs — OpenAI, Anthropic, Google DeepMind — pressed into the next model generation in 2023 and 2024, infrastructure costs multiplied again. Dario Amodei stated publicly in April 2024 that frontier model training was then costing approximately $100 million, that models already in development would cost approximately $1 billion, and that the trajectory pointed toward $5–10 billion per run by 2025–2026 [8]. Independent academic analysis confirmed the underlying trend: frontier model training costs had been scaling at roughly 2.4 times annually [9]. The cost curve continued upward; the capability differential over the best open source alternatives continued to compress. Each dollar of additional training expenditure was buying less advantage over open source competitors. The marginal return on pre-training investment had begun its structural decline before open source competition made the economics undeniable.

The pre-training data advantage claimed by frontier labs was itself largely illusory: foundation model training draws predominantly from publicly available internet text — web crawls, digitized books, code repositories — accessible to any team willing to process it. This data is not proprietary; it is the shared commons of the public internet. It is also legally contested. OpenAI stated in submissions to the UK government that training modern AI systems without copyrighted material would be "impossible" [10] — an acknowledgment of structural dependency, not a concession of liability; whether such training constitutes infringement remains actively litigated in multiple jurisdictions. The pre-training data commons is doubly vulnerable as a competitive position: it provides no proprietary competitive advantage, and no unambiguous legal title.

The data moat, to the extent one exists at all, lives not at the pre-training layer but in the proprietary fine-tuning and operational data accumulated through specific deployment relationships over time.

Open source made the underlying economics undeniable. Meta had established open-weight models as a serious frontier-capable alternative as early as February 2023, when LLaMA-1 demonstrated that a well-resourced lab could achieve competitive performance and release the weights publicly — a strategic commitment that predated both DeepSeek and OpenAI's gpt-oss by years [11]. Meta's Llama 3 family matched GPT-4 class performance for the majority of practical applications [12][13]. Then, in January 2025, DeepSeek dissolved the remaining argument. DeepSeek-R1, released by a Chinese AI laboratory, produced reasoning capabilities matching or exceeding OpenAI's o1 model at a reported training cost of under $6 million [14] — approximately GPT-3 era expenditure, for the class of reasoning capability that frontier labs had projected would require billions to produce. In April 2024, Dario Amodei had projected frontier training costs of $5–10 billion by 2025–2026 [8]. DeepSeek-R1 arrived at the low end of that projected window at roughly one-thousandth of the projected cost. The gap was not a rounding error. It was a structural refutation of the cost narrative on which the entire moat argument rested.

The generation-by-generation cost escalation was not abstract. GPT-2, released by OpenAI in 2019 with 1.5 billion parameters, cost approximately $43,000 to train — estimated from the published hardware parameters of 32 TPU v3 chips running approximately 168 hours [15][16]. GPT-3, released in June 2020 with 175 billion parameters — scaling from 1.5 billion to 175 billion — carried an estimated training cost of $4 to 5 million [5][16]. GPT-4, released in 2023, carried an estimated training cost of $60 to 100 million — an order-of-magnitude increase over GPT-3, achieved with an order-of-magnitude greater parameter count [6][16]. GPT-4.5, released in early 2025, carried training costs estimated at approximately $500 million — roughly five times GPT-4, a trajectory that appeared to validate the escalation narrative [17][18]. The slope was consistent and steep across four generations: from $43,000 to $500 million in six years — approximately four orders of magnitude. Each generation arrived with a training cost multiplier roughly proportional to its scale increase; the raw compute scaling rate across the frontier model population ran at an estimated four to five times per year, compounding the cost trajectory with every model generation [16].

The trajectory also had a destination that no one stated plainly in public, though the extrapolation was circulating seriously in private industry discussions as recently as 2024. If training costs scaled as projected — $100 million in 2024, $1 billion in 2025, $5–10 billion by 2026 — the logical endpoint by the early 2030s was training runs in the hundreds of billions of dollars. Extended further, the numbers approach a trillion. That extrapolation was never interrogated publicly because the narrative served everyone positioned inside it: higher projected training costs simultaneously elevated the apparent moat and justified the valuation of the next funding round. DeepSeek demonstrated the costs were not technically necessary — at production scale, in public, with a technical report attached [14][19].

What made DeepSeek's efficiency possible was not luck or exceptional resources. The underlying DeepSeek-V3 base model was trained at roughly 22 tokens per total parameter, nearly exactly the Chinchilla compute-optimal ratio of approximately 20 [4][19]. A mixture-of-experts architecture activated only 37 billion of 671 billion total parameters per forward pass, reducing effective compute per token by an order of magnitude. FP8 precision training further compressed hardware requirements throughout [19]. None of these choices are proprietary. All three are documented in public technical reports available to any research team. US frontier labs knew these techniques — they were not proprietary, and several originated from the US research community — but the hyperscaler circular economy that financed their operations created structural incentives to consume compute rather than minimize it: high training expenditure was simultaneously the mechanism of valuation inflation and the justification passed to the next investor in the chain. The pre-training fallacy was exposed: the moat was a pricing structure, not a technical floor, and disciplined engineering eliminated the pricing structure.

The economics of pre-training and inference are not symmetric, and understanding the distinction matters for what follows. Pre-training is a large, infrequent capital expenditure: a frontier model training run consumed millions of GPU-hours over weeks or months and was performed once per model generation. Inference is the ongoing operational expenditure: running a trained model to produce outputs at whatever scale commercial deployment demands, repeated billions of times across the model's life. These are different activities requiring different hardware. AWS provides architecturally distinct chips for each workload — Trainium for training, Inferentia for inference — reflecting the different computational demands of each. NVIDIA's training dominance, established through H100 and B200 GPU architectures, faces a different competitive landscape at inference, where throughput per dollar per token matters more than raw parallel compute. As pre-training commoditizes and model weights distribute freely, inference becomes the primary ongoing cost of AI deployment — and a different strategic battleground.

DeepSeek is not an isolated event. China's open source AI ecosystem has produced multiple frontier-capable model families across independent institutions: DeepSeek from High-Flyer Capital's research lab, Qwen from Alibaba [20], and GLM from Tsinghua University's Zhipu AI. These represent sustained, parallel development across the Chinese research landscape. As of early 2026, the leading open source reasoning models originate predominantly from Chinese institutions, are freely available to any actor globally, and were built at costs accessible to a wide range of state and non-state actors. The US government was watching.

What is underway is a market correction — not of AI's underlying capabilities, which are real and consequential, but of the assumptions about where AI value accrues, which proved to be badly wrong.

The deeper architectural case against the scale hypothesis is not only economic. A large language model is, at its foundation, a natural language interface: a system trained to process and generate human-language artifacts with statistical fluency. What scaling cannot produce is epistemic authority — the capacity to distinguish genuine knowledge from fluent confabulation. Models hallucinate because they have no mechanism to verify claims against grounded external reality; they generate plausible text, not warranted assertions. They become stale because training data has a cutoff date, the world continues, and retraining at frontier scale is prohibitively expensive — not a temporary engineering problem but a structural consequence of the pre-training paradigm. These are not bugs that additional parameters will fix. They are architectural properties of the approach, defining the limits of what pre-training alone can deliver. The solutions — retrieval augmentation to connect models to current verified knowledge, test-time reasoning with external verification, agentic tool use, multi-agent cross-checking — are not incremental improvements to the base paradigm. They are architectural extensions that address what scale structurally cannot.

The deeper shift — one that has received insufficient attention — is that pre-training scale has been decoupled from frontier AI capability. Three developments arrived in parallel, each partially severing the link between parameter count and practical performance. Post-training optimization, through reinforcement learning from human feedback, instruction fine-tuning, and preference learning, delivers substantial practical capability improvements at a fraction of pre-training cost; much of what users experience as model capability reflects post-training work, not raw scale.[2] Test-time compute scaling, the architectural insight underlying both DeepSeek-R1 and OpenAI's o1, relocates the compute budget from training to inference, allowing models to reason step by step before producing a final answer — meaning a smaller, cheaper model with sufficient inference budget can outperform a larger, more expensive one on reasoning tasks. And agentic composition replaces the singleton model as the unit of AI capability entirely: rather than a single large dense model handling all tasks end-to-end, capable systems are built by composing foundation models with retrieval systems, tool-use frameworks, and orchestration layers for multi-step planning, with mixture-of-experts architectures applying the same principle internally by activating only task-relevant parameters per query. GPT-5, released in August 2025, made the decoupling concrete: it was trained on less pre-training compute than GPT-4.5 — the first reversal in a decade-long trajectory of per-generation scaling increases — not because OpenAI reduced ambition but because scaled post-training on a smaller base model outperformed a larger one, with the capability differential residing in the post-training work rather than the pre-training run [18]. Whether future models return to larger pre-training runs is analytically secondary: the existence of the GPT-5 pathway proves that frontier capability no longer requires massive pre-training spend.[3] The link has been decoupled, not merely deferred — and a decoupled moat is no moat at all.

Intelligence, in the engineering sense, is increasingly a property of the system, not the model.

The singleton — one large, closed, proprietary model as the terminal destination of AI capability — was both a market strategy and a technical claim. The technical claim has not held.

## The Hyperscaler Circular Economy

Understanding why foundation model valuations reached the levels they did — and why those valuations are now structurally compromised — requires understanding a specific financial arrangement that has gone underanalyzed.

The mechanism, stated plainly: a hyperscaler writes an equity check into a foundation model company at a valuation it declares — and simultaneously collects revenue when that company spends the capital on compute, at the high margins scaled cloud infrastructure commands. The declared valuation then becomes the floor that subsequent investors anchor to: each successive round prices from the previous round's declared value, not from independently verified fundamentals. The hyperscaler does not need to participate in later rounds to benefit from this dynamic — its equity appreciates as each anchored round inflates the asset upward, while its infrastructure revenue compounds with every dollar of compute the company is structurally required to spend. A conventional venture investor pressures portfolio companies to minimize infrastructure burn. When the investor and the infrastructure vendor are the same party, that discipline disappears entirely: high compute spend is simultaneously desirable as a revenue stream and as justification for the next, higher valuation. Escalating training costs were not a technological inevitability — they were the engine of the valuation, and valuation was the product being sold to the next investor in the chain.

The AI industry organized into two identifiable blocs: Microsoft and its Azure cloud anchoring OpenAI; Amazon Web Services and Google Cloud jointly anchoring Anthropic. These are not passive financial relationships. Microsoft has invested, across successive funding rounds, a cumulative sum widely reported at approximately $13 billion [21]. Amazon committed an initial $4 billion to Anthropic in September 2023 [22], with a second $4 billion tranche in November 2024 bringing total commitment to $8 billion and formalizing AWS as Anthropic's "primary cloud and training partner" [23]. Google has contributed additional capital as a cloud and equity partner. In operational terms, both OpenAI and Anthropic function as the flagship AI products of competing cloud infrastructure empires. Their compute dependency on their hyperscaler partners is not a business risk to be managed — it is an existential structural condition.

These investments created a self-reinforcing economic loop. In each cycle, the hyperscaler commits capital to the foundation model company. The foundation model company spends that capital on cloud compute — predominantly from the same hyperscaler's infrastructure. The compute spend flows back to the hyperscaler as revenue. Escalating pre-training costs, enabled by the investment, justify a higher valuation at the next round. The next round repeats the cycle at greater scale. The hyperscalers earned returns on both sides of each transaction: equity appreciation in the foundation model company, and high-margin revenue on the infrastructure the company was required to purchase — revenue generated at the economics of scaled cloud, where each additional dollar of workload spend contributes disproportionately to operating income. High pre-training costs, far from being an unavoidable technical constraint, were the mechanism that made the structure work — creating the appearance of a capital-intensive moat that suppressed competition while justifying each successive valuation increase.

The Amazon-Anthropic relationship makes the structure concrete in every detail. Amazon invested $8 billion across two tranches [22][23]. Anthropic's financial relationship with Amazon operates on two distinct tracks: a commitment to purchase Trainium chips, a capital expenditure flowing to Amazon's silicon business as hardware revenue; and ongoing cloud services for training workloads, a recurring operational expense that generates revenue for AWS. Both streams flow to the same investor that set the round price and holds the equity stake [23][24]. By February 2026, Amazon's stake in Anthropic had appreciated to $60.6 billion in reported value — a seven-fold return on $8 billion [24]. Amazon classified this stake as a "Level 3" asset: financial accounting terminology for a valuation based not on observable market prices but on the company's own internal models and what SEC disclosures call "unobservable inputs" [24]. The valuation rests on assumptions made by the same party that set the purchase price.

The Anthropic valuation history renders the structure in quantitative terms. From approximately $1 billion at founding in 2021, the company's implied valuation reached $4.1 billion by April 2022, $15 billion following the first Amazon investment in September 2023, and $18.5 billion by February 2024 [25]. Then the acceleration that tracked Amazon's deepening involvement: a $3.5 billion Series E raise in March 2025 at a $61.5 billion post-money valuation [25]; a $13 billion Series F raise in September 2025 at $183 billion [26]; and a $30 billion Series G raise in February 2026 at $380 billion [27]. From $1 billion to $380 billion in approximately five years — a 380-fold increase — with the most explosive appreciation occurring in the period when Amazon was simultaneously the largest investor, the primary infrastructure vendor, and the party responsible for marking the convertible notes to value using its own internal assumptions. Critically, the $183 billion and $380 billion valuations were reached after DeepSeek had already demonstrated that the cost trajectory underlying the moat narrative was false. The circular economy continued running even after the mechanism was exposed.[4]

In a private funding round, the valuation is not discovered through competitive price discovery — it is declared by whoever leads the round. When the lead investor is also the company's primary vendor, capturing high-margin revenue on every dollar the company subsequently spends on compute, the valuation it sets serves its interests simultaneously as equity appreciation and as justification for continued, high-cost compute spend flowing back to its own infrastructure. The headline figure — "foundation model company now valued at $X billion" — is less a market signal than a coordinated price that creates the conditions for the next round. Each round passes the asset to a buyer who accepts the previous buyer's price as the floor, not the ceiling. This is the structure of a hot potato: it passes from party to party as long as there is a next buyer willing to receive it. The circular economy was still running at full acceleration in March 2025 — more than two months after DeepSeek had publicly demonstrated that the cost floor underpinning the moat narrative was false — when OpenAI closed a $40 billion Series F round led by SoftBank, the largest private technology funding round ever recorded, valuing OpenAI at $300 billion [28].

The terminal event is the IPO, which passes the asset to public markets where independent buyers reprice it on fundamentals. That terminal buyer — retail investors, pension funds, and institutional capital — will price the asset on the economics that the private round structure was not designed to produce. Both OpenAI and Anthropic have been accelerating toward public listings, which represent the necessary exit for prior investors at the valuations those investors declared. Unlike a private round chain, there is no next buyer after the public market. The hot potato lands.

The structure served both parties as a form of rent extraction. Hyperscalers charged above-competitive rates for compute — rates insulated from competitive discipline by the dependency relationship embedded in the equity investment. If a foundation model company must spend its investor's capital on that investor's infrastructure, the price discovery that would otherwise discipline compute pricing does not function. Foundation model companies, in turn, aspired to extract rent from their closed APIs: if general intelligence was available only through one or two endpoints, those companies could charge prices that reflected quasi-monopoly positioning rather than marginal production costs. Open source dissolved the API rent before it could be collected at scale. DeepSeek dissolved the infrastructure cost illusion. Both forms of rent extraction depended on the same false premise: that pre-training at frontier scale was prohibitively expensive and technically irreproducible. It was neither.

DeepSeek demonstrated, at production scale, that frontier reasoning capability did not require hyperscaler-priced compute. The apparent moat was a sunk cost, underwritten by infrastructure margins. The circular economy that justified it lost its foundation.

Foundation model companies face the commodity trap in its classical form: prices driven toward marginal cost by open source competition, while fixed costs — infrastructure, talent, the compute spend itself — remain enormous. The billions spent on pre-training did not create defensible advantages. They created liabilities. Neither OpenAI nor Anthropic is profitable: both companies generate substantial revenues — Anthropic's annualized revenue run-rate reached approximately $3 billion by May 2025 and approximately $9 billion by end of 2025 [29] — but both continue to require large external capital infusions to fund operations in which infrastructure costs consistently exceed revenues. OpenAI generated approximately $4.3 billion in revenue in the first half of 2025 alone, against a cash burn of approximately $2.5 billion in the same period [30], with internal projections suggesting cumulative losses could reach $115 billion through 2029 [31]. Revenue has grown substantially; losses have grown faster. The costs are not too high because the companies lack customers. The costs are too high because hyperscaler infrastructure pricing — set by the same parties who invested in these companies and marked their valuations — consumes revenues at rates that no pre-training moat can justify once that moat is exposed as illusory.

## Where Value Has Migrated

Value does not reside in models. Models are interchangeable inference engines, increasingly available at negligible marginal cost — from open source providers at near-zero cost, or from commercial APIs at prices approaching that floor. The AI industry is discovering, at speed, what utility economics could have predicted: the company that builds essential infrastructure does not necessarily capture the value flowing through it. Early electric utilities generated all the economic activity that electricity enabled — and earned the regulated margin on transmission. The railroad companies that built the physical infrastructure of the industrial economy saw the steel producers, grain traders, and manufacturers who used the rails accumulate the wealth. Foundation model companies built the cognitive infrastructure of the AI economy. The value that flows through that infrastructure will be enormous. The question is who captures it — and the evidence increasingly points away from the infrastructure providers themselves.

The first dimension of value is proprietary knowledge and structured data. In the commercial domain, this means a healthcare system with thirty years of longitudinal patient records, a financial institution with proprietary market microstructure data, or a law firm whose accumulated case outcomes represent knowledge that no internet crawl can replicate. In the national security domain, it means knowledge graphs of adversary logistics networks, operational planning records, and classified intelligence assessments. Foundation model capabilities are general. This knowledge is specific, accumulated over time, and specificity is where value lives.

The second is operational deployment infrastructure. Integrating AI into consequential workflows — financial risk management, clinical decision support, military logistics, intelligence synthesis — requires security frameworks, compliance architecture, institutional approval processes, and the organizational trust that comes only through demonstrated reliability over time. Foundation model companies provide APIs. Operational deployment is a different and more durable capability set.

The third is personnel and clearance infrastructure. Access to classified applications is not primarily a technical question. It is a personnel question. Government work at the classified level requires security-cleared teams, and clearances are granted to individuals on the basis of rigorous background investigation — not to companies on the basis of stated values or published safety frameworks. The commercial AI ecosystem, built substantially on internationally diverse talent, operates under personnel standards structurally incompatible with classified environments. This mismatch cannot be resolved through policy commitments.

The fourth is trust accumulated through operation. In both commercial and national security contexts, the vendors who will hold the strongest positions are those who have demonstrated — through sustained deployment in consequential environments — the security culture, operational reliability, and institutional accountability that high-stakes work requires. A company that has run AI reliably in financial trading for five years has something a company that launched a chatbot last year does not. That gap widens over time and cannot be purchased.

The most durable moats in the post-foundation-model economy will not be model quality — which commoditizes — but the accumulated operational knowledge, workflow integration, and institutional trust that specific deployments generate over time. The post-foundation-model value stack is clear: commodity inference at the base; integration, execution, proprietary data, and operational trust at the top.

The asymmetry between hyperscaler durability and foundation model fragility has a structural explanation that the circular economy obscured. Cloud infrastructure creates three forms of lock-in that model APIs cannot replicate: data gravity — moving petabytes between cloud providers is expensive, slow, and disruptive in ways that switching an API endpoint is not; integration depth — enterprises embed cloud infrastructure at every architectural layer, from networking and storage to identity and compliance, over years of accumulated dependency; and regulatory moat — cloud providers have obtained FedRAMP, IL5, and classified environment authorizations that required years and substantial investment to achieve, creating genuine barriers to entry unrelated to technical quality. Model weights can be copied in minutes. API endpoints can be replaced in an afternoon. Training methodology is documented in public papers. There is no data gravity, no integration depth, and no regulatory moat at the foundation model layer. This is why the hyperscaler extracted durable rent from the circular economy while the foundation model company could not — and why the structure that appeared to benefit both parties was always asymmetric in favor of one.

## III. The National Security Arbiter

### A Predictable Assertion of Control

The relationship between strategic technology and government authority does not follow a single arc — but it arrives at the same destination. Nuclear technology was developed under direct government control from its inception: the Manhattan Project was a wartime government program, and private nuclear capability has since been permitted only under stringent regulatory frameworks, with no private actor authorized to develop nuclear weapons. The internet emerged from DARPA-funded academic research before commercial actors built the public network — and the government today operates SIPRNET and JWICS as classified networks physically and logically separate from the commercial internet. Space launch was government-defined before commercial participants entered under FAA licensing frameworks, with government as the dominant anchor customer for early commercial ventures. Telecommunications were privately built, then regulated, then subjected to mandatory government access requirements for national security purposes. Cryptography followed the same arc — developed commercially, classified as a munition under ITAR, and subjected to mandatory key escrow demands before the technology distributed beyond regulatory reach. The classified internet and the commercial internet developed in parallel across all of these domains: no traffic crosses between them, they are governed by different rules, accessible to different populations, and optimized for different workloads. AI is following the same structural path — not by deliberate design, but because the same underlying logic that demanded classified networks demands a classified AI development track.

Artificial intelligence belongs in this genealogy. Its foundational research — the Dartmouth Conference of 1956, the neural network and natural language processing work of subsequent decades, the machine learning advances that preceded the current commercial era — was substantially funded by DARPA grants, NSF programs, and public universities. Like nuclear physics and packet-switched networking before it, AI was a twentieth-century public research investment whose strategic implications are being reaffirmed, seventy years later, as the hardware finally caught up. The paths to government control differ across these technologies. The destination does not. Every dual-use technology that materially alters the calculus of state power has been brought under government authority in national security applications — regardless of whether its origins were public or private, and regardless of how its developers characterized their intentions. AI is not exempt from this logic. It could not be.

The proliferation dynamics differ across these technologies — AI model weights can be distributed globally within hours in ways that nuclear capability cannot — but the government's assertion of authority follows the same destination: control over strategic applications of dual-use capabilities, not ownership of the underlying physics or mathematics.

The National Security Commission on Artificial Intelligence concluded in its 2021 Final Report [32] that AI is a strategic technology requiring government oversight, investment, and control in national security applications. That report was not speculative. It was a policy statement of what was coming.

The only genuinely surprising element is that foundation model companies believed otherwise — that the combination of technical sophistication, safety-focused branding, and investor backing would exempt them from the accountability standards applied to every other vendor of strategic national security technology.

The logic of the national security AI market has a categorical quality that commercial market participants consistently underappreciate: the government does not participate in this market as a buyer among others. It constitutes the market — by determining, through the clearance and vetting process, which vendors are authorized to operate within it at all. There is no regulatory workaround, no premium service tier that substitutes for clearance, and no technical capability that bypasses the pre-authorization that classified work requires.

That market is not a direct bilateral relationship between the government and its AI vendors. It is tiered. Authorized defense contractors and system integrators — the cleared companies that build operational systems for government clients — sit between the government's end-use requirements and the foundation model capabilities they deploy. These contractors hold clearances, develop mission-critical systems, and are themselves authorized vendors of the state. A foundation model company embedded as upstream infrastructure for those contractors' operational systems becomes part of a supply chain whose reliability the government has a direct interest in protecting. When a foundation model vendor introduces volatility — through service interruptions, governance disputes, or unilateral policy changes — the disruption does not stop at the vendor relationship. It propagates through every authorized contractor that has built operational capability on that infrastructure. 10 U.S.C. § 3252 is the statutory instrument for addressing precisely this: an unreliable component anywhere in a military supply chain creates risk for every system and operation downstream.

The urgency of that supply chain authority has been sharpened by a specific competitive dynamic: Chinese institutions have developed sustained frontier AI capability and distributed it freely and globally — through open-weight releases available to any actor, including contractors and integrators within US defense supply chains — outside the procurement and vetting frameworks the US government applies to its domestic vendors.

China has demonstrated sustained capacity to develop frontier AI systems with state backing, operating under different data access rules and facing no reciprocal export controls on its own development. The asymmetry is significant but layered: US hardware companies face export restrictions on their highest-capability chips — the October 2023 and October 2024 Commerce Department rules constraining NVIDIA H100, H200, and B200 class GPU sales to China and designated countries[5] — while US foundation model companies face the inverse vulnerability: open-source import competition from Chinese institutions whose models distribute freely and globally without equivalent constraints. These are different companies facing different vulnerabilities at different layers of the stack. The export controls constrain what China can build at the hardware layer; they do not constrain Chinese open-weight models from competing globally at the software layer. For the domestic vendors on whom the US defense supply chain depends, this competitive dynamic makes AI supply chain provenance and reliability a live operational concern rather than a theoretical one. The national security establishment cannot maintain sensitive applications on infrastructure built by companies whose personnel, security practices, and economic stability have not been vetted to the standards that classified work requires.

The standard being applied is not new. It is the one applied to every vendor of strategic national security technology. AI is not exempt.

## The Anthropic Case

The immediate sequence that produced Anthropic's designation matters less than the structural incompatibility it exposed — but the sequence is worth establishing precisely so that the structural argument is not mistaken for a political one.

The proximate trigger was a specific operational episode. In January 2026, Claude was reportedly accessed via Palantir's defense AI platform during a US military operation in Venezuela [33][34]. When this was leaked to journalists in mid-February, it crystallized a conflict that had nothing to do with the merits of that particular operation and everything to do with a categorical governance incompatibility: the Pentagon demanded that Anthropic accept terms permitting "all lawful use" of Claude, while Anthropic maintained two specific restrictions it regarded as non-negotiable — prohibitions on use for lethal autonomous warfare and mass surveillance of Americans [2]. According to Anthropic's complaint, it had largely complied with the government's broader demands — the impasse, in Anthropic's account, was over those two carve-outs. What the dispute exposed was not a disagreement about the merits of autonomous weapons policy — that is a question for governments and democracies, not vendors. It was a question of authority: whether a private company retains the right to withhold service from the state for uses the state has determined are lawful. The government gave Anthropic a public ultimatum with a 5:01pm deadline on February 27: accept "all lawful use" or face either designation as a supply chain risk under 10 U.S.C. § 3252 or compelled compliance under the Defense Production Act — the emergency procurement statute previously used to commandeer private production capacity for wartime and national emergency requirements. The DPA threat made explicit what the structural analysis in this paper describes as the inevitable destination: the government's assertion that it does not recognize a private company's right to maintain a "usage policy" that overrides determinations of lawfulness made by the state. That is an assertion of sovereign authority, and the same standard applied to Huawei, Kaspersky, and every prior vendor of strategic national security technology.

Notably, Claude was already being deployed in AWS GovCloud and AWS Secret and Top Secret Cloud Regions as of the November 2024 Amazon partnership announcement [23] — the government's objection was not to classified cloud deployment per se, but to the governance terms under which that deployment occurred and the financial stability of the vendor providing it.

The conflict the designation addressed was not limited to the government's own direct use of Claude — it extended to the reliability of the authorized contractor ecosystem that had built operational systems on that capability.

The conflict escalated through late February. A leaked internal Anthropic memo critical of the administration — for the tone of which CEO Dario Amodei publicly apologized — compounded the political dimension in the days before the decisive moment [1][35]. The memo's critical tone confirmed what the governance conflict had already signaled: not merely policy disagreement but a cultural posture at odds with the institutional disposition that trusted vendor status requires. On February 27, 2026, the President ordered federal agencies to cease use of Anthropic's AI systems, and the Secretary of War designated Anthropic a supply chain risk effective immediately [36][37] — under both DoD supply chain management authority (10 U.S.C. § 3252) and federal ICT supply chain security frameworks (41 U.S.C. § 1323) [38][39][40], the same statutory authorities previously applied to Huawei and Kaspersky. The parallel is instructive: AI models are now treated as critical supply chain infrastructure requiring the same provenance and control standards applied to telecommunications hardware and security software. The government's response extended beyond contract termination: new federal procurement rules in development would require AI vendors to grant irrevocable rights for lawful government use — resolving the underlying governance conflict structurally rather than case by case [41][42].

The Anthropic case reflects three structural failures that are analytically distinct.

The first is a failure of political economy. Anthropic's strategy rested on a premise that was always untenable: that a private company could define the ethical terms of AI use for the US government, maintain those terms as binding conditions on government contractors, and position that authority as a virtue rather than a liability. The complaint itself confirms the precise shape of this failure: the remaining dispute, in Anthropic's account, was over two specific carve-outs. The government's position was not that Anthropic's safety views were wrong. It was that Anthropic did not have the authority to impose those views as conditions on the state's use of a strategic technology. That is a question of sovereign authority. The distinction matters: the government does not participate in the national security technology market as a buyer that can be persuaded by better pricing, stronger safety credentials, or superior capability. It is a sovereign authority that determines which vendors are authorized to participate at all — a categorically different relationship than any commercial procurement. Constitutional AI — Anthropic's proprietary framework for governing model behavior through a closed, non-auditable internal ruleset — is a coherent instrument for commercial product differentiation. It is also load-bearing for the business model: the usage policy is the enforcement mechanism of the closed model, which is the source of the API moat, which is the basis for the commercial premium over open-weight alternatives. The government's demand for irrevocable use rights was therefore not merely a request to change terms of service — it was a structural demand that Anthropic relinquish the governance architecture on which its commercial valuation rests. As a governance framework for a strategic national security technology, it has a categorical deficiency: it is the vendor's unilateral assertion of what is permissible, enforced through model weights that no external party can inspect or verify. Safety claimed through a closed system is not safety in any institutional sense that a sovereign government deploying that system can accept. It is control exercised by the vendor, not by the state.

The second is an economic disqualifier independent of the governance conflict. Anthropic's operations are sustained by continuous large funding rounds from Amazon and Google — the same hyperscalers whose compute infrastructure Anthropic depends upon. A vendor requiring perpetual external capital infusions to remain operational is a systemic risk for mission-critical applications regardless of technical capability. Government procurement standards have always required vendor stability. AI is not exempt from that requirement, and no amount of technical capability or stated mission offsets the instability of a company that cannot operate without its next funding round. The economic fragility is not unique to Anthropic: it is structural to the foundation model business model itself. In November 2025, OpenAI's CFO Sarah Friar floated the idea of federal loan guarantees to backstop AI infrastructure financing, describing government support as a mechanism to lower borrowing costs for the industry [31]. The suggestion was immediately criticized and subsequently retracted: CEO Sam Altman — who had himself testified before the Senate Commerce Committee in May 2025 on US AI competitiveness and infrastructure investment needs [43] — stated publicly that "we do not have or want government guarantees for OpenAI datacenters" and that "taxpayers should not bail out companies that make bad business decisions" [31]. The episode is a precise indicator of the structural condition: a company whose CFO reaches for federal backstops as a financing tool, and whose CEO must publicly disavow that reach while simultaneously lobbying Congress for supportive AI infrastructure policy, is not operating from a position of strength. The government's public designation of Anthropic cited governance conduct, not financial instability — and the same economic disqualifier applies with comparable force to OpenAI, whose CFO's loan guarantee suggestion reflects the same structural condition. The government's selective enforcement reflects a deliberate competition strategy: maintaining viable vendors across a range of integration postures, with financial fragility as a structural backdrop rather than a stated disqualification criterion. Financial fragility is this paper's structural analysis of a latent supply chain risk endemic to the foundation model business model — equally applicable to OpenAI as to Anthropic, and operative as a background condition rather than the government's stated rationale for any specific designation. The government's response reflects a competition strategy that manages this structural risk by maintaining multiple vendors at different levels of integration.

The third structural disqualifier operates independently of both governance philosophy and financial stability: the personnel incompatibility between commercial AI development and classified work. Security clearances in the US national security context are granted to individuals following rigorous background investigation — not to companies based on stated values or published frameworks. The commercial AI ecosystem, built substantially on internationally mobile talent with connections across research institutions in multiple countries, operates under personnel norms and mobility patterns that classified environments cannot accommodate. This incompatibility is not a policy problem that commitment or investment can resolve. It would require years of deliberate organizational restructuring — hiring specifically for clearance-eligible populations, rebuilding research culture around operational security requirements — and no foundation model company has undertaken that process.

Anthropic's public response has substituted a values debate for the accountability question actually at issue. Framing the designation as political persecution, and positioning Constitutional AI as evidence of unique trustworthiness, addresses neither operational sovereignty nor financial stability. The structural gap between government relevance and actual organizational design is precise and complete: the prerequisites — an auditable open-weight model, clearance-compatible security infrastructure, demonstrated financial stability, and a personnel base compatible with classification requirements — are not what a commercially optimized frontier AI company is built to provide, and not what Anthropic has built. There is a further structural irony: a closed-weight model whose internal logic cannot be independently audited is itself a supply chain risk for any sovereign deployer. The government cannot verify what it cannot inspect. A model whose behavior is governed by a proprietary internal ruleset with no external audit pathway does not reduce dependency — it relocates it, from hardware and infrastructure to the vendor's unilateral judgment about what the model will and will not do.

The deeper regulatory principle at work is one the government applies to every utility and common carrier in its supply chain: an infrastructure provider embedded in mission-critical operations cannot maintain discretionary authority over service delivery to lawful users. Electric utilities cannot interrupt power to defense contractors because they disapprove of what is being manufactured. Telecommunications carriers cannot selectively deny service based on ideological objection to end-users. The Defense Production Act threat — the government's explicit alternative to the supply chain designation — operationalized this principle directly: when private infrastructure becomes essential to defense production, the state retains authority to commandeer its use, removing vendor discretion entirely. The procurement requirement for "irrevocable government use rights" encodes the same principle in peacetime statutory form. The government's demand was not primarily about restriction content — it was a structural demand that the vendor relinquish discretionary authority over service delivery to lawful users. A closed-model governance architecture — proprietary weights, unauditable behavior, a vendor retaining unilateral authority to modify or withdraw service — introduces supply chain risk that persists independent of what the current usage policy says.

The structural logic points toward what a government facing these disqualifiers actually needs: not a better-governed commercial foundation model company, but sovereign access to open-weight models that cannot be withdrawn by vendor policy, that can be fine-tuned on classified data in air-gapped environments with no vendor in the loop, and that give the government direct control of the weights for modification and deployment at will. Open-weight models solve the governance incompatibility at its source — they have no usage policy to override, no vendor financial dependency to manage, and no clearance mismatch to resolve through years of organizational restructuring. The irrevocable government use rights now required by emerging federal procurement standards [42] are, in structural terms, a demand for open-weight access through policy where the market has not provided it organically. The section that follows examines how one foundation model company has begun moving toward that architecture while another has not.

## The Pattern

The Anthropic designation is not an isolated incident [44]. It is the pattern asserting itself. Strategic technologies move from private innovation to government gatekeeping. The timing is determined by capability thresholds, security incidents, and the arrival of foreign parity — all of which converged in 2025 and 2026. The outcome was structurally determined long before it arrived.

On March 9, 2026, Anthropic filed suit in US District Court (Northern District of California), Case No. 3:26-cv-01996-RFL, naming as defendants not only the Department of War but seventeen federal departments and agencies — Treasury, GSA, State, HHS, Commerce, Veterans Affairs, DHS, SEC, NASA, Energy, the Federal Reserve, and others — reflecting the government-wide scope of the presidential directive [45][46]. The complaint, which documents the 5:01pm ultimatum and the Defense Production Act threat in detail, advances five legal claims: that the designation was arbitrary and capricious under the Administrative Procedure Act (no reasoned explanation, no evidence, no consideration of less restrictive alternatives); that it constituted First Amendment retaliation for Anthropic's protected speech about AI safety and its refusal to accept certain terms of service; that it violated the First Amendment Petition Clause by punishing Anthropic for lobbying and policy advocacy; that it denied Fifth Amendment due process (no notice, no hearing, no opportunity to respond before contract cancellations began); and that all downstream agency actions implementing the designation are independently unlawful under the APA. The suit is argued to be the first federal procurement exclusion of a US AI company under these authorities — a claim that, if the court accepts jurisdiction, would make this a significant administrative law precedent regardless of outcome. The structural analysis in this paper is independent of the litigation's resolution. If the designation is vacated on procedural grounds, the three structural incompatibilities identified above — governance incompatibility, financial fragility as a background structural condition, and personnel mismatch — remain fully intact and will produce the same confrontation on the next occasion. If upheld, the precedent extends the authority to the full range of AI vendors operating in national security contexts. The pattern does not depend on this particular ruling.

New federal AI procurement standards requiring irrevocable government use rights signal policy direction [42]. The institutional posture is confirmed in Anthropic's own account of its interactions with the Department of War [1]. The government is not regulating AI from the outside. It is selecting which vendors participate in building the AI systems it will depend upon, and establishing the terms on which that participation is possible.

**The Parallel Navigation**

The Anthropic case is instructive not only for what it reveals about that company's structural position but for what the contrast with OpenAI's simultaneous navigation of the same period reveals about the structure itself.

OpenAI's posture followed a different sequence entirely. In August 2025, OpenAI released two open-weight frontier models — gpt-oss-120b and gpt-oss-20b — under Apache 2.0 license, its first open-weight release since GPT-2, simultaneously making both models available for direct download on Hugging Face and through Microsoft's Azure AI Foundry [47][48]. OpenAI made the governance tradeoff explicit in its own language. On adversarial risk, OpenAI stated directly that it had assessed the threat by fine-tuning the model on sensitive domain data to simulate how an attacker might proceed — "creating a domain-specific non-refusing version for each domain the way an attacker might" — and released the models anyway, having determined the strategic benefits outweighed the proliferation risk [47]. On the geopolitical rationale, OpenAI wrote: "Broad access to these capable open-weights models created in the US helps expand democratic AI rails" [47]. The governance tradeoff was not implicit. OpenAI named it, assessed it, and chose. A government agency in an air-gapped classified environment can deploy, fine-tune, and operate gpt-oss with no OpenAI relationship required.

In October 2025, months before the Anthropic-Pentagon conflict became public, OpenAI submitted a formal policy proposal to the White House Office of Science and Technology Policy recommending sovereign AI infrastructure for the US government — specifically, architecture enabling the government to operate frontier AI "outside of hyperscaler constraints while providing the trusted, high-performance infrastructure essential for the next generation of defense and intelligence capabilities" [49]. OpenAI was proposing the institutional structure that the government's procurement standards were already demanding, before the conflict that made those demands public.

On February 28, 2026, while the Anthropic designation was escalating, OpenAI announced a formal agreement with the Department of War [50]. The agreement retained OpenAI's safety stack and specified three explicit red lines: no use for mass domestic surveillance, no direction of autonomous weapons systems, no high-stakes automated social credit decisions. Those red lines — no autonomous weapons direction, no mass surveillance — were substantively identical to the two restrictions Anthropic had maintained as non-negotiable. The government accepted those terms from OpenAI the day after rejecting them from Anthropic. The dispute was not about what the restrictions said. It was about which vendor the government was willing to authorize. Anthropic's public narrative — that it was penalized for safety principles the government wanted overridden — is contradicted by the parallel deal. The restrictions were acceptable. The company was not. More precisely: a vendor that retains discretionary authority to modify or withdraw service from mission-critical infrastructure — regardless of what its current policy says — is structurally incompatible with reliable supply chain deployment. The risk is not the restriction. It is the discretion. The agreement was amended in early March to add an explicit restriction: DoW intelligence agencies, including the NSA, would not use OpenAI's commercial API [51]. The NSA does not conduct sensitive intelligence work through commercial APIs — the amendment describes existing operational reality, not a new constraint on government use. The open-weight gpt-oss models, deployed eight months earlier, had already resolved the governance problem at its source: the government can operate them in air-gapped classified environments with no OpenAI relationship and no usage policy to negotiate. The February 28 agreement was not assembled in response to the Anthropic designation. It was the conclusion of a process that had been running in parallel for months — the open-weight release in August 2025, the sovereign AI infrastructure proposal to the White House in October 2025, and the formal agreement in February 2026 were a single coherent sequence, not a reactive improvisation. The governance conflict that forced Anthropic's designation does not arise when the government controls the weights. The open-weight release is not merely a commercial strategy. It is the technical mechanism by which frontier model capability can be adapted, fine-tuned on classified data, and deployed on the national security track without commercial vendor involvement — the architecture Section VI identifies as the two-track path's operational foundation.

The contrast is structural, not a matter of character or trustworthiness. OpenAI adapted its governance architecture — open-weight release, sovereign AI proposal, direct agreement with explicit red lines — to the same forces that Anthropic met with policy enforcement and legal challenge. Neither company has fully resolved the foundational tension between commercial optimization and classified national security requirements — but one has created a structural path toward sovereign deployment; the other has not.

The government, for its part, is not simply selecting a winner. A government that excluded its only capable AI vendor would have no capability. A government with only one capable AI vendor would have no leverage. The simultaneous Anthropic exclusion and OpenAI agreement reflects a deliberate competition strategy: maintaining vendors at different levels of integration, retaining the credible threat of exclusion as an accountability mechanism, and ensuring that no single foundation model company achieves the quasi-monopoly position the foundation model era promised. The AI governance market, like other strategic technology markets before it, is not winner-take-all from the government's perspective. Competition itself is a governance tool — and the government is deploying it.

## IV. The Singleton Fallacy

The foundation model era was organized around a structural misconception: that one company would build AGI — understood here as systems capable of reliably planning and executing complex tasks across domains — control it through scale and first-mover advantage, and the rest of the world would access intelligence through its API. This singleton capture model assumed winner-take-all dynamics in which the largest training run produced the dominant system, competitors were priced out by capital requirements, and the winning company achieved a monopolistic position in the supply of artificial general intelligence.

The commercial posture of the leading foundation model companies was explicit on this point, even when not stated in those terms. The strategic logic of both OpenAI and Anthropic rested on the premise that whoever achieved AGI first would define and control it — that the prize was not merely market share but something closer to permanent epistemic authority over the most consequential technology ever developed. The corollary was that this authority would be self-conferred: the winning company would declare itself the responsible steward, set the terms of access, and govern the technology's deployment through its own policies rather than through external accountability structures. Neither governments nor competitors nor the broader public were cast as legitimate counterparties in this governance — only the winning lab and the investors and personnel who shared its mission.

This was not a strategic error of execution. It was a foundational error of political economy. No private company has ever been permitted to govern a genuinely dual-use technology on its own terms — and AGI is dual-use by definition. The belief that this one would be the exception required ignoring not just every historical precedent but the self-evident nature of the capability being built. The pattern is nuclear, space, telecommunications, cryptography — always the same destination. Cryptography is the closest precedent: the government ultimately lost the Crypto Wars as open-source encryption distributed beyond regulatory reach, but it established the institutional principle — sovereign access to the capability, even when held by private parties, is non-negotiable — before conceding the practical fight. Open-weight models are, in this sense, the key escrow the government is now demanding by other means: sovereign access to the weights, regardless of who built them. In every case, private actors who believed the pattern did not apply to them were wrong. The foundation model companies were not uniquely positioned to be exceptions to it. The belief that they were explains most of what has gone wrong for them in 2026.

The assumption failed simultaneously on all four dimensions it required to hold: that pre-training scale produced a durable moat (it did not); that training cost constituted a prohibitive barrier to entry (it was a cost structure enforced by circular financing, not a genuine technical ceiling, as DeepSeek demonstrated); that governments would remain passive observers of private AGI development (never plausible, now explicitly contradicted by statute and policy); and that general intelligence would emerge from a single large dense model trained end-to-end (superseded by the evidence that intelligence is a system property, not a model property, with the paradigm having shifted decisively from "train the model larger" to "compose the system better"). Any one of these failures would have restructured the industry. All four arrived together.

What has replaced the singleton model is more distributed, more competitive, and more structurally stable. Multiple models at rough performance parity compete for inference workloads. Open source ensures continuous improvement outside any single company's control. The application and integration layer is where differentiation and value are accumulating. There is no winner-take-all outcome at the model layer because the model layer has commoditized. The value has moved up the stack, and it will not move back.

## V. The Post-Foundation-Model Industry

### The Value Stack, Inverted

With value having migrated away from the model layer, the AI market is splitting into segments with fundamentally different requirements, trust models, and economics — and the segments are not permeable. The most consequential division is between commercial AI and national security AI: not merely different markets, but architecturally incompatible ones. The division is also an inversion along the open/closed axis. Commercial AI is deployed openly — public-facing, publicly benchmarked, globally distributed — but depends on closed weights to maintain commercial value: the proprietary model architecture is the source of the API moat, the usage policy, and the commercial premium. National security AI is deployed in closed, classified environments — air-gapped, sovereign, non-public — but requires open weights to maintain operational sovereignty: without inspectable, government-controlled weights, every deployment creates vendor dependency, potential service interruption, and unverifiable behavior in active operations. The open-weight model resolves both requirements simultaneously. It dissolves the commercial moat — and enables precisely the closed sovereign deployment that national security operations require.

Commercial AI — serving consumer and enterprise applications — is large, competitive, and commoditizing rapidly at the infrastructure layer. The competitive field includes two structurally distinct categories: OpenAI and Anthropic are pure foundation model companies whose entire commercial strategy depends on pre-training large language models and monetizing API access, while Google, Meta, and xAI are conglomerates that produce foundation models as one business among many — and for whom the foundation model position is not existential. Google's Gemini family, for instance, is one capability among many within a search, advertising, and cloud business that does not require Gemini to win the model race to remain structurally sound. xAI's March 2025 acquisition of X Corp. in an all-stock transaction illustrates the first form this logic takes: Grok's model capabilities absorbed into a social distribution platform of hundreds of millions of users, converting the AI capability from a standalone product into a retention and engagement instrument [52]. The SpaceX acquisition of xAI in February 2026 illustrates a structurally different move — not application-layer integration but defense-industrial absorption. SpaceX completed the acquisition, making xAI a subsidiary of a company with existing Space Force and Air Force contracts, Starlink deployed as strategic communications infrastructure, and clearance relationships built through years of government launch business [53][54]. xAI is no longer a foundation model company in any meaningful sense: it is a capability embedded in the defense-industrial base, now structured by the institutional relationships of its parent rather than by the commercial API economics of its origin.

Meta has uniquely committed to open-weight releases since LLaMA-1 in February 2023, contributing more to the open source AI ecosystem than any other major lab and predating both DeepSeek and gpt-oss as in establishing open-weight models as a frontier-capable alternative to closed APIs [11] — a commitment sustained through successive Llama generations and undiminished under competitive pressure. OpenAI, Anthropic, Google, and the open source ecosystem — distributed primarily through platforms like Hugging Face, which hosts hundreds of capable models available at near-zero marginal cost — are the primary competitors in this segment. Inference cost is the primary variable. Differentiation comes from application-layer integration, not model quality.

National security AI — serving defense, intelligence, and government applications — operates under entirely different constraints. Security clearances, vetted personnel, air-gapped deployment environments, and demonstrated operational reliability are prerequisites, not preferences. Foundation model companies that cannot meet these requirements are excluded from this market. The exclusion is structural and permanent, not a temporary regulatory condition awaiting resolution.

A company optimized for one segment cannot easily reposition for the other. The personnel infrastructure, security architecture, and institutional relationships required for national security work cannot be purchased. The commercial AI ecosystem and the national security AI ecosystem are diverging, and the divergence is structural.

That divergence is already visible in the behavior of sophisticated commercial actors. Microsoft — which has invested approximately $13 billion in OpenAI across successive funding rounds [21] — has simultaneously made a strategic $16 million investment in Mistral

AI, a French open-weight model company whose European backing was driven explicitly by sovereign AI concerns about closed-model concentration, and which reached a $14 billion valuation in September 2025 as ASML led a €1.7 billion round [55][56][57]; integrated Meta's open-weight Llama models into its product suite; and developed its own internal Phi model family. The European sovereign AI concern — that governments cannot tolerate strategic dependency on foreign-controlled closed-weight models whose weights cannot be inspected, whose behavior cannot be audited, and whose API access can be withdrawn at the vendor's discretion — anticipated the US government's 2026 confrontation with Anthropic by several years. Germany's Aleph Alpha built its entire model architecture around mandatory data residency and weight auditability for government clients, because European sovereigns demanded exactly those terms. What Mistral and Aleph Alpha represented as a European demand in 2023 and 2024, the US government operationalized through procurement law in 2026. Microsoft's simultaneous hedging of its OpenAI position reflects the same logic applied commercially: when the largest investor in a foundation model company actively builds and distributes alternatives to its product, the market has rendered its verdict on whether foundation models constitute a durable moat. They do not. Foundation models have become infrastructure, and every sophisticated operator of infrastructure ensures it does not depend on a single supplier. OpenAI itself released two open-weight frontier models under Apache 2.0 license in August 2025 — gpt-oss-120b and gpt-oss-20b — its first open-weight release since GPT-2, a strategic acknowledgment that frontier capability can no longer be held exclusively behind a closed API [47].

## The Commercial Track's Own Correction

The national security bifurcation is the most structurally definitive dimension of the current shift, but the commercial AI market is undergoing its own correction — not from government intervention but from the economic logic of commoditization itself. The irrational exuberance of the foundation model era — marked by messianic claims about AGI singularities, predictions of the end of software engineering as a profession, and valuations premised on quasi-monopoly control of general intelligence — is correcting toward something closer to utility economics. Foundation model companies now face the trajectory that early utilities, railroad companies, and first-generation telecommunications providers followed: they built transformative infrastructure, proved the concept at enormous cost, and discovered that the infrastructure converged toward commodity pricing while value accumulated in the hands of those who used it.

What is clear in March 2026 that was not clear in 2022 is that there was never going to be a multi-trillion-dollar arbitrage on artificial general intelligence. The terminal market structure for cognitive infrastructure looks more like the electricity grid than like a monopoly search engine: multiple providers, near-commodity pricing, value captured by users rather than infrastructure operators. The companies that will hold durable structural positions in the AI economy — positions that translate into significant valuations twenty-five years from now — are more likely to be those that embed AI capability so deeply into proprietary operational data and irreplaceable workflows that the model underneath becomes interchangeable while the accumulated knowledge system built on top of it becomes the moat. Many of those companies may not yet exist in recognizable form, or exist today as narrow-domain players whose strategic position is not yet apparent from the outside — much as Andrew Carnegie was not visible as the future consolidator of American steel from the vantage of 1870, when the railroads were the obvious winners of the industrial revolution and Carnegie's mills were merely customers of the infrastructure. Carnegie's ascent came in the 1880s and 1890s, not through building railroads but through disciplined use of Bessemer steel technology and vertical integration across a supply chain the railroads made possible — until he was ultimately acquired by J.P. Morgan in 1901 and folded into U.S. Steel. Carnegie himself was not the enduring winner; he was the first great user of the infrastructure who was then consolidated.

Standard Oil offers an even sharper illustration: Rockefeller's dominance was built in part on negotiating preferential rail rates from railroad companies desperate for volume, turning the railroad infrastructure into a competitive lever that destroyed rivals in a different industry entirely. The railroads needed Standard Oil's freight more than Standard Oil needed any single railroad. The enduring value accrued not to the infrastructure builders but to those who figured out how to use it asymmetrically — and the companies positioned to do that in the AI economy are not yet obvious. This trajectory — innovative technology commoditizing under competitive pressure while value migrates to the users of infrastructure rather than its providers — is the pattern documented across the innovation economics literature [58][59][60].

The behavioral signals are already accumulating from multiple directions. OpenAI — the company most closely identified with the closed-API singleton strategy — released frontier-capable open-weight models in August 2025 [47] and, in October 2025, submitted a formal policy proposal to the White House recommending sovereign AI infrastructure that would enable the government to operate frontier AI "outside of hyperscaler constraints" [49]. A company that once organized its mission around controlling access to frontier intelligence through a closed API has begun simultaneously distributing model weights freely and proposing that the US government operate AI infrastructure it controls. The correction is not only economic and technical. It is institutional.

The foundation model companies with the largest capital positions face a specific version of this question: whether a well-funded incumbent can execute the application-layer pivot before the commodity trap fully forecloses it. The structural argument of this paper does not require them to fail as companies — only that the foundation model position itself is not where durable value resides, regardless of capitalization or execution quality.

## The NVIDIA Position

One structural beneficiary of foundation model commoditization deserves explicit recognition. As model weights distribute freely and inference prices collapse, the hardware layer captures the value the software layer cannot retain. H100s and B200s remain essential infrastructure for training and inference across commercial, open source, and government systems alike.

NVIDIA's strategic posture confirms this logic. Jensen Huang stated publicly that NVIDIA would not invest $100 billion in OpenAI and was unlikely to take a large equity position in any single AI company [61]. The reasoning is structurally sound: NVIDIA earns more by selling compute infrastructure to every player in the ecosystem — commercial labs, open source developers, government contractors — than by concentrating equity exposure in any one of them. When the model commoditizes, the pick-and-shovel supplier wins. That NVIDIA itself declines to bet on which model company prevails is the market's clearest signal that none of them has a durable moat.

NVIDIA's posture is further illuminated by its December 2025 transaction with Groq. Reuters reported, citing CNBC, a transaction valued at approximately $20 billion — potentially NVIDIA's largest-ever deal — involving a license to Groq's inference technology and the hiring of key Groq personnel who joined NVIDIA as part of the agreement [62][63]. Groq's own announcement described a non-exclusive inference technology licensing agreement [63]. The precise deal structure has been characterized variously across reporting. Jensen Huang stated publicly that "we are not acquiring Groq as a company," describing instead a technology licensing and talent transfer arrangement: founder Jonathan Ross and President Sunny Madra joined NVIDIA, while a new CEO (Simon Edwards) leads the continuing Groq entity, which operates its GroqCloud inference platform independently. The reported $20 billion figure derives from CNBC reporting rather than disclosed financials. What the transaction signals, regardless of its precise structure, is unambiguous: NVIDIA recognizes that inference is the next competitive battleground. Groq's inference chip architecture — built as a Language Processing Unit (LPU) purpose-designed for transformer inference throughput — represents a fundamentally different design philosophy from the GPU architecture optimized for the massively parallel compute required during training. As training commoditizes and model weights distribute freely, inference economics — throughput per dollar per token across billions of daily queries — become the primary variable in AI hardware competition. A reported $20 billion to secure access to specialized inference architecture reflects a clear-eyed

judgment about where the hardware value is moving.

## VI. The Two-Track Path to AGI

### From Model to System

AGI will not emerge from pre-training scale alone. This is now clear on both technical and economic grounds. What it will emerge from — what it is, in specific operational domains, already beginning to resemble — is the integration of foundation model capabilities with proprietary knowledge, agentic execution, and real-world deployment feedback.

The technical frontier has moved accordingly. The important unsolved problems are not in scaling pre-training runs. They are in building systems that reason reliably across multiple steps, integrate with live knowledge bases without hallucination, take consequential actions safely in complex environments, and maintain coherent behavior across extended operational workflows. Retrieval-augmented generation (RAG) [64], agentic reasoning frameworks [65], and multi-agent coordination [66] represent the active research frontier. The governance and trust infrastructure for enterprise agent deployment — standardized capability verification, permission management, and identity frameworks for agents operating across organizational boundaries — represents a parallel and largely unsolved challenge [67]. These are system-level problems, not model-scale problems.

The commercial AI ecosystem is converging on this architecture from below: application companies are building agentic systems that compose foundation model outputs with retrieval systems, tool-use frameworks, and workflow orchestration. The foundation model companies themselves are offering agentic APIs and multi-modal capabilities in recognition that the base model alone is no longer the product. The question is whether the companies that built the base models can hold durable positions in the systems built on top of them — or whether the integration layer becomes the seat of value entirely, with foundation models as interchangeable infrastructure underneath.

### The Permanent Divergence

The most important structural development in AI that current public discourse has not adequately registered: commercial AGI and national security AGI are bifurcating onto separate development trajectories — a structural divergence that, once established, does not reverse.

The commercial track is built on publicly available training data, open benchmarks, published research, and academic collaboration. Its models are evaluated against public datasets and deployed in civilian contexts. Its safety research is peer-reviewed. Its capability advances are visible in demo videos and user reports. This track will produce genuinely capable systems — agentic, multimodal, increasingly integrated into consequential workflows. But it is not the most capable track.

The national security track is built on something qualitatively different. Classified intelligence databases — imagery, signals, human intelligence, communication intercepts — represent decades of collection that no commercial training set can replicate. Operational planning records encode decision-making patterns, adversary behavior, and strategic context that have no commercial analog. The systems being built on this data, in classified environments, evaluated against real operational requirements, and deployed in active intelligence and military contexts, are not constrained by the commercial training pipeline or the public research discourse. They are not on public benchmarks. Their capabilities are not announced in press releases. The open-weight releases now entering the market — trained on public data, distributed freely, fine-tunable without vendor involvement — represent the technical pathway by which commercial-origin model capability enters this environment: weights that can be adapted on classified corpora in air-gapped systems, with no commercial vendor in the loop and no usage policy to override.

That divergence is already visible in the behavior of sophisticated commercial actors. The Maven Smart System — the DoW AI platform at the center of the Anthropic conflict — was used for intelligence analysis in active strike operations [68], including during the US campaign in Iran, in which the system integrating Claude suggested and prioritized targets during an opening phase that struck roughly one thousand targets within the first twenty-four hours [69]. The government's 5:01pm ultimatum on February 27 — and its invocation of the Defense Production Act as the alternative instrument of compulsion — was issued on the eve of that operation's escalation. The governance ambiguity was not a theoretical concern about future use cases. It was an active operational constraint in a live conflict. The same model company simultaneously facing government exclusion was operationally embedded in the systems the government depended upon. This is not contradiction. It is the transition moment made visible: the government knows it needs this capability, and it is in the process of establishing the terms on which it will access it — on its terms, not the company's.

The SpaceX acquisition of xAI illustrates a third mechanism by which the national security track is forming — distinct from vendor exclusion and distinct from open-weight adoption. SpaceX did not navigate the clearance and accountability requirements that blocked Anthropic: it already satisfied them, through Space Force launch contracts, Air Force relationships, and Starlink's role as deployed military communications infrastructure, all predating the foundation model era. Its financial stability is not contingent on perpetual venture capital infusions. Its personnel security infrastructure was not built for AI — it was built for aerospace and defense, and the AI capability has been acquired into it rather than the reverse. The acquisition places Grok inside an entity that can deploy frontier AI in national security contexts without a vendor relationship with a capital-dependent foundation model company, without usage policies subject to governance conflicts, and without the clearance mismatch that made Anthropic structurally incompatible. This is the national security track asserting itself not through government procurement but through the defense-industrial base acquiring AI capability directly [53][54]. The directional logic runs both ways: SpaceX brings Grok into the defense-industrial base, and simultaneously brings defense-grade infrastructure — Starlink's satellite communications network, already deployed as strategic military communications — into a platform serving hundreds of millions of consumer users through X. AI capability flows into defense; defense infrastructure flows into commercial and consumer deployment. The traditional boundary between the two was never absolute, but it is dissolving further.

The most consequential AI development of the next decade will occur on the national security track. It will be invisible to public discourse. It will be trained on data the commercial ecosystem cannot access, evaluated against operational requirements that public benchmarks do not capture, and deployed in contexts that press releases do not describe. The public AI ecosystem — commercial foundation models, open source alternatives, academic research — is the visible surface of a much larger structure. The iceberg below is classified.

## VII. Strategic Implications

The structural shift described in this paper is not a prediction. It has already occurred. Its implications for capital allocation, company strategy, and policy are consequential and immediate.

For those allocating capital, the core reorientation is this: foundation model companies face structural compression, not temporary headwinds. The circular economy that produced their valuations has been dissolved by open source parity. The moats they described were sunk costs underwritten by hyperscaler infrastructure margins. The application and integration layer — companies with proprietary data, operational deployment capability, and where relevant, security clearance infrastructure — is where value is accumulating. The government AI market is structurally underappreciated: US total national defense spending exceeded $1 trillion in FY2026 [70], with a presidential proposal to reach $1.5 trillion by FY2027 [71], and AI-augmented operations in that context represent a large addressable market with fewer competitors, higher margins, and stickier relationships than the commercial AI space. NVIDIA holds the clearest hardware moat in the ecosystem, confirmed by its own posture of selling compute to all players rather than concentrating equity in any one, and by its reported $20

billion commitment to secure the inference hardware market before specialized competitors could establish it.

For companies building in AI, the strategic imperative follows from the same analysis: stop competing on model quality and build moats that models cannot commoditize. Proprietary knowledge bases, agentic execution systems embedded in operational workflows, security and compliance infrastructure, and the institutional trust that comes from demonstrated reliability in consequential environments — these are the assets that compound. Architecture should be model-agnostic; the foundation model is interchangeable infrastructure, and dependence on any single provider is a negotiating liability. The defining technical transition of this period is the shift from chat interfaces to agentic execution — systems that plan, act, and adapt across multi-step workflows — and the companies that internalize this earliest will hold the structural position that matters.

For policymakers, the Anthropic designation is not a conclusion but a precedent. The government is actively selecting which vendors participate in building national security AI infrastructure, applying the same operational security and stability standards it applies across all strategic technology procurement. Government strategy documents from 2023 and 2024 confirm the institutional posture across agencies [72][73][74]. The emerging framework requiring irrevocable government use rights for AI vendors [42] resolves the underlying governance conflict the Anthropic case exposed. Open-weight models are not the opposite of sovereign control — they are its primary instrument. A government that holds the weights commands the capability on its own terms, without vendor intermediation. The counterintuitive logic of this claim deserves explicit statement: the same government that might reflexively frame open-weight AI as a proliferation risk has the strongest structural incentive to ensure that capable open-weight models exist, are domestically developed, and remain available for sovereign deployment — because a government that controls the weights controls the capability, without depending on any vendor's policy, financial continuity, or personnel clearance.

A government that cannot depend on capital-dependent vendors with their own governance terms, non-cleared personnel, and externally marked valuations has an affirmative structural incentive to ensure that capable open-weight models exist at frontier performance levels, are domestically developed and auditable by government personnel, and are available for sovereign deployment in air-gapped classified environments. Open-weight models solve the governance conflict the Anthropic case exposed: no vendor can revoke them through a policy change or a contract termination, they do not create financial dependency on companies requiring perpetual capital infusions, and they permit the government to fine-tune, modify, and deploy without vendor involvement. The gpt-oss release demonstrates that even the company most identified with closed-API control has begun moving toward open weights under commercial and strategic pressure [47] — the market is converging on what national security requirements have already demanded. Maintaining a competitive domestic open-weight ecosystem at the frontier is not a commercial policy preference. It is a national security imperative for any government whose adversaries are developing the same dual-use capabilities without equivalent constraints.

Maintaining multiple AI vendors at different levels of integration — with exclusion as a credible and exercised accountability mechanism — is not an artifact of the Anthropic conflict but a replicable governance strategy for any technology market where capability is concentrated and vendor conduct cannot otherwise be verified. The classified development track requires commensurate investment: cleared contractors and in-house government AI capabilities built without dependency on foundation model companies requiring perpetual external financing. SpaceX's acquisition of xAI is the first publicly documented case of this architecture assembling itself through private-sector initiative rather than government procurement — a defense contractor with pre-existing clearance infrastructure acquiring frontier AI capability and integrating it with the institutional relationships that commercial foundation model companies have not been able to provide.

For researchers, the marginal return on pre-training scale work has diminished materially relative to the open problems in agentic reasoning, reliable multi-step planning, knowledge integration, and robust deployment in operational environments. Security clearances are increasing in research value. The national security AI track offers access to unique data, real operational constraints, and feedback from high-stakes deployment — a research environment qualitatively different from public benchmark competition.

## VIII. A Note on Scope, Evidence, and Version

This paper covers a story whose principal events were still unfolding at the time of writing. Version 1.0 documents the structural shift in AI through March 9, 2026. Readers encountering this paper later will know more than was knowable on that date: the Anthropic litigation will have advanced, resolved, or been appealed, the federal AI procurement standards requiring irrevocable government use rights will have been finalized or contested, the OpenAI-Department of War agreement will have been tested against operational reality, and the two-track divergence described in Section VI will be more or less visible depending on what has been declassified. The structural arguments are designed to be durable regardless of how those specific developments resolve. The factual record they rest on is the record as of March 9, 2026. Subsequent versions of this paper will incorporate material developments in the litigation, procurement standards, and the national security AI track as that record becomes public.

The government actions described became public between February 13 and early March 2026; the operative designation — the Secretary of War's supply chain risk determination, effective immediately — occurred February 27, 2026, with Anthropic's legal challenge filed March 9. Primary source documentation is noted in the bibliography with URLs where confirmed. The Anthropic valuation data draws on multiple financial press sources across rounds, each cited directly: CNBC for the March 2025 Series E round [25], Reuters for the September 2025 Series F [26], and The Guardian for the February 2026 round [27], anchored by the Business Insider disclosure of Amazon's stake valuation and Level 3 accounting classification [24]. Revenue figures for both Anthropic and OpenAI derive from journalistic reporting citing industry sources; as private companies, neither publishes audited financial statements. The paper's claims about the national security AI track are structurally inferred from the logic of classified data access and operational deployment, and from publicly reported operational deployments — the classified track is, by definition, not fully documented in public sources. "AGI" is used throughout in its operational sense: systems that reliably plan and execute complex tasks across domains, not a hypothetical threshold event.

## IX. Conclusion: The Correction

The foundation model era ends not with a dramatic failure but with a structural correction — the kind that follows every sustained period of irrational exuberance in transformative technology. The messianic rhetoric of 2022–2024 — AGI by a specific year, the end of software engineering as a profession, economic singularities just ahead — was the symptom of a market organized around assumptions that could not be tested until they were tested. They have now been tested. They did not hold. The correction does not mean AI is less important than its proponents claimed. It means that where AI value accrues is entirely different from where the foundation model era located it.

The simultaneous falsification of three beliefs organized billions of dollars of capital and years of institutional ambition.

The first belief — what this paper calls the scaling fallacy: that pre-training scale compounds indefinitely into a durable moat. It does not. The capital poured into pre-training runs was not wasted — it pushed the frontier of AI capability forward, demonstrated that general-purpose language intelligence could be engineered at scale, and produced the open source infrastructure that is now freely available to the world. But it was never recoverable as a private advantage. DeepSeek and its successors demonstrated that the cost barrier was a pricing structure, not a technical floor, and that anyone with the engineering discipline to train compute-optimally could replicate frontier results. The sunk cost of pre-training has been absorbed into the commons. It was, in the end, a public good.

The second belief: that the government would remain a passive observer of private AGI development. It would not, and it has not. The NSCAI said so in 2021. Every precedent from nuclear to space, communications infrastructure, and cryptography pointed the same direction. The Anthropic designation confirmed it under statute. The new procurement rules requiring irrevocable government use rights have now encoded it as policy. A private company cannot define the terms of lawful use for the US military, and cannot reserve the right to withdraw AI capability from active operations based on its own assessment of appropriate use. The government's assertion of control is not persecution. It is the rational enforcement of the same standards applied to every other vendor of strategic technology.

The third belief: that AGI would emerge at the foundation model layer and be controlled by whoever built the largest model. It will not. AGI — in the operationally meaningful sense — emerges from integrated systems: foundation models combined with proprietary knowledge, agentic execution, real-world feedback, and the trust infrastructure that makes consequential deployment possible. No single commercial AI company will control it. The most capable systems are already being built on the national security track, invisible to public benchmarks, trained on data that no commercial actor can access — and the institutional structure of that track is already being assembled. SpaceX's acquisition of xAI is one visible data point in a process that is otherwise largely classified: the defense-industrial base acquiring frontier AI capability and integrating it with the clearance infrastructure, operational relationships, and financial stability that commercial foundation model companies have not been able to provide.

What the structural logic demands is not a better-governed commercial vendor. It is sovereign access to open-weight frontier models that the government can audit, fine-tune, and deploy on its own terms — without usage policies that override operational requirements, without financial fragility that creates supply chain risk, and without vendor relationships that cannot survive the clearance and accountability standards that classified work demands. That demand is already encoded in emerging procurement standards requiring irrevocable government use rights. It is already driving the market's most significant behavioral signal — the company most identified with closed-API control releasing open-weight frontier models under maximally permissive licensing. And it is already constituting the entry architecture for the national security AI track this paper identifies as the most consequential development of the coming decade. The structural solution is not coming. It is already being deployed.

What comes after the foundation model era is clearer than the hype cycle suggested. Inference is infrastructure. Execution is the product. The value is in integration, data, trust, and the institutional relationships that make consequential deployment possible. The most consequential AI development is already underway on a track that public discourse cannot see.

The structural choices being made now — about which vendors access which data, which architectures receive investment, and which companies are trusted to build the systems that will matter most — will define the shape of AI capabilities for the decade ahead. The foundation model era is over. What comes next is already being built.

The structural argument this paper makes is not a prediction. It is a description of what has already happened — and a map of what follows from it.

## Endnotes

[1] "Department of War" refers to the secondary title authorized by Executive Order 14347 (September 5, 2025), which restored the historical name as an authorized title for the Department of Defense. The statutory name remains the Department of Defense; the executive order authorizes use of "Department of War" in official correspondence, contracts, and operational contexts [75]. This paper uses "Department of War" and "DoW" throughout, consistent with the usage adopted in official government and partner documentation during the period covered.

[2] The significance of post-training optimization as a driver of practical AI capability — distinct from raw pre-training scale — is documented in the RLHF literature originating with Ziegler et al. (2019) and developed through the InstructGPT work of Ouyang et al. (2022). The central insight: a model optimized through human feedback to be helpful, harmless, and honest can substantially outperform a larger model trained purely on next-token prediction, for most practical applications users actually care about. The capability that end-users experience is largely a post-training artifact, not a raw scale artifact.

[3] The Epoch AI analysis underlying this claim projects that GPT-6 will likely return to larger pre-training compute runs as post-training scaling approaches its near-term ceiling [18]. This projection does not affect the structural argument. The claim is not that OpenAI will never scale pre-training again — it is that frontier capability has been demonstrated without massive pre-training spend, which is sufficient to refute the cost-barrier moat narrative. A moat defined by prohibitive cost is dissolved the moment the cost is shown to be unnecessary, regardless of what subsequent models do. GPT-5 answered the question. The answer is: large pre-training runs are not required for frontier capability. That answer does not expire when GPT-6 arrives.

[4] A counterargument holds that late-stage valuations reflect not the continuation of the pre-training moat narrative but a pivot thesis — that investors were pricing Anthropic as a future application-layer or safety-layer winner. If so, the expected multiple would reflect application-layer software economics: high-growth SaaS companies typically trade at 10–20x forward revenue. Against Anthropic's annualized revenue run-rate of approximately $3 billion by May 2025 [29] — the period proximate to the September 2025 Series F raise — a $183 billion valuation implies a 61x multiple. Against a $9 billion run-rate by end of 2025 [29], the February 2026 valuation of $380 billion implies 42x. These are not application-layer multiples. They are pre-training moat multiples — pricing continued scarcity and monopoly control of frontier intelligence, not future integration revenue. The circular economy continued running on its original assumptions even as those assumptions were being falsified in public. The same structural condition applies to OpenAI: its CFO's November 2025 suggestion of federal loan guarantees reflects comparable capital dependency. The government's decision to contract OpenAI while excluding Anthropic reflects competition strategy and institutional trust, not a differential financial assessment between two comparably fragile vendors.

[5] The October 2023 Bureau of Industry and Security (BIS) interim final rule and the October 2024 expansion established export licensing requirements for advanced AI accelerators above defined computational thresholds — specifically targeting NVIDIA H100, H200, A100, and B200 class GPUs — for export to China and a range of designated countries. The rules also established requirements for US cloud providers serving foreign customers with compute above those thresholds. Chinese AI laboratories, including

DeepSeek, have publicly reported use of older NVIDIA A100 clusters acquired before controls took effect and domestically produced alternatives including Huawei Ascend chips. The controls constrained but did not eliminate Chinese frontier AI development. The asymmetry the paper describes is not that China has unconstrained access to advanced compute — it does not — but that China faces no equivalent restrictions on data collection at population scale, no equivalent transparency requirements for model development, and no reciprocal export controls limiting distribution of open-weight models developed on constrained hardware. The asymmetry is further layered by industry type: US hardware companies (NVIDIA, AMD) face export restrictions on outbound chip sales; US foundation model companies (OpenAI, Anthropic) face import competition from Chinese open-source models that distribute freely globally. These are different companies at different stack layers facing crossed, not parallel, vulnerabilities. The US does not restrict export of open-weight models — Meta's Llama and OpenAI's gpt-oss are available globally — creating symmetric model-layer competition while the hardware layer remains asymmetrically constrained. The practical result is that China cannot easily acquire the newest training chips but can access and build upon open-weight models trained on those chips by US labs.

---

## Bibliography

[1] Amodei, D. (2026, March 5). Where things stand with the Department of War. *Anthropic Newsroom*. https://www.anthropic.com/news/where-stand-department-war

[2] O'Brien, M. (2026, March 6). Pentagon's chief tech officer says he clashed with AI company Anthropic over autonomous warfare. *AP News*. https://apnews.com/article/pentagon-anthropic-ai-autonomous-warfare-emil-michael

[3] Kaplan, J., McCandlish, S., Henighan, T., Brown, T. B., Chess, B., Child, R., Gray, S., Radford, A., Wu, J., & Amodei, D. (2020). Scaling laws for neural language models. *arXiv:2001.08361*.

[4] Hoffmann, J., Borgeaud, S., Mensch, A., Buchatskaya, E., Cai, T., Rutherford, E., de las Casas, D., Hendricks, L. A., Welbl, J., Clark, A., Hennigan, T., Noland, E., Millican, A., van den Driessche, G., Damoc, B., Guy, A., Osindero, S., Simonyan, K., Elsen, E., Rae, J. W., Vinyals, O., & Sifre, L. (2022). Training compute-optimal large language models. *arXiv:2203.15556*.

[5] Brown, T. B., Mann, B., Ryder, N., Subbiah, M., Kaplan, J., Dhariwal, P., Neelakantan, A., Shyam, P., Sastry, G., Askell, A., Agarwal, S., Herbert-Voss, A., Krueger, G., Henighan, T., Child, R., Ramesh, A., Ziegler, D. M., Wu, J., Winter, C., Hesse, C., Chen, M., Sigler, E., Litwin, M., Gray, S., Chess, B., Clark, J., Berner, C., McCandlish, S., Radford, A., Sutskever, I., & Amodei, D. (2020). Language models are few-shot learners. *Advances in Neural Information Processing Systems, 33*. arXiv:2005.14165.

[6] OpenAI. (2023). GPT-4 technical report. *arXiv:2303.08774*.

[7] Epoch AI. (2024). Training compute costs are doubling every eight months for the largest AI models. *Epoch AI Data Insights*. https://epoch.ai/data-insights/cost-trend-large-scale

[8] Amodei, D. (2024, April 23). *CNBC Squawk Box* [interview transcript]. CNBC. https://www.cnbc.com/2024/04/23/cnbc-exclusive-cnbc-transcript-anthropic-co-founder-ceo-dario-amodei-speaks-with-cnbcs-andrew-ross-sorkin-on-squawk-box-today.html

[9] Cottier, B., Rahman, R., Fattorini, L., Maslej, N., & Owen, D. (2024). The rising costs of training frontier AI models. *arXiv:2405.21015*.

[10] OpenAI. (2023). Response to the UK's copyright consultation. https://openai.com/global-affairs/response-to-uk-copyright-consultation/

[11] Touvron, H., Lavril, T., Izacard, G., Martinet, X., Lachaux, M.-A., Lacroix, T., Rozière, B., Goyal, N., Hambro, E., Azhar, F., Rodriguez, A., Joulin, A., Grave, E., & Lample, G. (2023). LLaMA: Open and efficient foundation language models. *arXiv:2302.13971*.

[12] Dubey, A., Jauhri, A., Pandey, A., Kadian, A., Al-Dahle, A., Letman, A., Mathur, A., Schelten, A., Yang, A., Fan, A., & et al. (2024). The Llama 3 herd of models. *arXiv:2407.21783*.

[13] Touvron, H., Martin, L., Stone, K., Albert, P., Almahairi, A., Babaei, Y., Bashlykov, N., Batra, S., Bhargava, P., Bhosale, S., & et al. (2023). Llama 2: Open foundation and fine-tuned chat models. *arXiv:2307.09288*.

[14] Guo, D., Yang, D., Zhang, H., Song, J., Zhang, R., Xu, R., Zhu, Q., Ma, S., Wang, P., Bi, X., & et al. (2025). DeepSeek-R1: Incentivizing reasoning capability in LLMs via reinforcement learning. *arXiv:2501.12948*.

[15] Radford, A., Wu, J., Child, R., Luan, D., Amodei, D., & Sutskever, I. (2019). Language models are unsupervised multitask learners. *OpenAI Blog*. https://cdn.openai.com/better-language-models/languag e_models_are_unsupervised_multitask_learners.pdf

[16] Epoch AI. (2026). AI models dataset. https://epoch.ai/data/large-scale-ai-models

[17] Seetharaman, D. (2024, December 20). OpenAI's next big AI effort, GPT-5, is behind schedule and expensive. *The Wall Street Journal*. https://www.wsj.com/tech/ai/openai-gpt5-orion-delays-639e7693

[18] Edelman, Y., Denain, J.-S., Sevilla, J., & Ho, A. (2025, September 26). Why GPT-5 used less training compute than GPT-4.5 (but GPT-6 probably won't). *Epoch AI Gradient Updates*. https://epoch.ai/gradi ent-updates/why-gpt5-used-less-training-compute-than-gpt45-but-gpt6-probably-wont

[19] DeepSeek AI. (2024). DeepSeek-V3 technical report. *arXiv:2412.19437*.

[20] Qwen Team. (2024). Qwen2.5 technical report. *arXiv:2412.15115*.

[21] Microsoft Corporation. (2024). *Form 10-K annual report*. U.S. SEC EDGAR. https://www.sec.gov/edgar/browse/?CIK=789019

[22] Anthropic. (2023, September 25). Expanding access to safer AI with Amazon. https://www.anthropic.com/news/anthropic-amazon

[23] Anthropic. (2024, November 22). Powering the next generation of AI development with AWS. https://www.anthropic.com/news/anthropic-amazon-trainium

[24] Kim, E. (2026, February 6). Amazon's $8 billion Anthropic investment balloons to $61 billion. *Business Insider*. https://www.businessinsider.com/amazon-ai-bet-anthropic-soars-61-billion-valuation-2026-2

[25] Capoot, A. (2025, March 3). Amazon-backed AI firm Anthropic valued at $61.5 billion after latest round. *CNBC*. https://www.cnbc.com/2025/03/03/amazon-backed-ai-firm-anthropic-valued-at-61point5-billio n-after-latest-round.html

[26] Reuters. (2025a, September 2). Anthropic's valuation more than doubles to $183 billion after $13 billion fundraise. https://www.reuters.com/business/anthropics-valuation-more-than-doubles-183-billion-after- 13-billion-fundraise-2025-09-02/

[27] The Guardian. (2026, February 12). Anthropic raises $30bn in latest round, valuing Claude bot maker at $380bn. https://www.theguardian.com/technology/2026/feb/12/anthropic-funding-round

[28] Field, H., & Rooney, K. (2025, March 31). OpenAI closes $40 billion funding round, the largest private fundraise in history. *CNBC*. https://www.cnbc.com/2025/03/31/openai-closes-40-billion-in-funding-the -largest-private-fundraise-in-history-softbank-chatgpt.html

[29] Tong, A., & Dastin, J. (2025, May 30). Anthropic hits $3 billion in annualized revenue on business demand for AI. *Reuters*. https://www.reuters.com/business/anthropic-hits-3-billion-annualized-revenue -business-demand-ai-2025-05-30/

[30] The Information. (2025). OpenAI's first half results: $4.3 billion in sales, $2.5 billion cash burn. https://www.theinformation.com/articles/openais-first-half-results-4-3-billion-sales-2-5-billion-cash-burn

[31] Reuters. (2025b, November 6). OpenAI discussed government loan guarantees for chip plants, not data centers, Altman says. https://www.reuters.com/business/openai-does-not-want-government-guarantees-massive-ai-data-center-buildout-ceo-2025-11-06/

[32] National Security Commission on Artificial Intelligence (NSCAI). (2021). *Final report*. https://reports.nscai.gov/final-report/

[33] Lawler, D., & Curi, M. (2026, February 13). Pentagon's use of Claude during Maduro raid sparks Anthropic feud. *Axios*. https://www.axios.com/2026/02/13/anthropic-claude-maduro-raid-pentagon

[34] Reuters. (2026a, February 13). US used Anthropic's Claude during the Venezuela raid, WSJ reports. https://www.reuters.com/world/americas/us-used-anthropics-claude-during-the-venezuela-raid-wsj-reports-2026-02-13/

[35] Business Insider. (2026, March 6). Anthropic CEO Dario Amodei apologized for the 'tone' of a leaked internal message criticizing the Trump administration. https://www.businessinsider.com/anthropic-ceo-dario-amodei-apologized-leaked-memo-criticizing-trump-administration-2026-3

[36] Anthropic. (2026, February 27). Statement on comments by the Secretary of War. *Anthropic*. https://www.anthropic.com/news/statement-comments-secretary-war

[37] Goodwin Procter LLP. (2026, March 5). Is Claude a supply chain risk? https://www.goodwinlaw.com/en/insights/publications/2026/03/alerts-practices-is-claude-a-supply-chain-risk

[38] Just Security. (2026). What Hegseth's "Supply Chain Risk" designation of Anthropic does and doesn't mean. https://www.justsecurity.org/132851/anthropic-supply-chain-risk-designation/

[39] 10 U.S.C. § 3252. Supply Chain Risk Management.

[40] 41 U.S.C. § 1323, enacted as Subtitle A of the Federal Acquisition Supply Chain Security Act of 2018 (the SECURE Technology Act, Pub. L. No. 115-390).

[41] Defense News. (2026, March 6). Pentagon says it is labeling Anthropic a supply chain risk 'effective immediately.' https://www.defensenews.com/news/pentagon-congress/2026/03/06/pentagon-says-it-is-labeling-anthropic-a-supply-chain-risk-effective-immediately/

[42] Reuters. (2026b, March 7). US draws up strict new AI guidelines amid Anthropic clash. https://www.reuters.com/business/media-telecom/us-draws-up-strict-new-ai-guidelines-amid-anthropic-clash-ft-reports-2026-03-07/

[43] U.S. Senate Committee on Commerce, Science, and Transportation. (2025, May 8). *Winning the AI race: Strengthening U.S. capabilities in computing and innovation* [hearing record]. https://www.commerce.senate.gov/2025/5/winning-the-ai-race-strengthening-u-s-capabilities-in-computing-and-innovation_2

[44] Bellan, R. (2026, March 5). Anthropic to challenge DoD's supply-chain label in court. *TechCrunch*. https://techcrunch.com/2026/03/05/anthropic-to-challenge-dods-supply-chain-label-in-court/

[45] Anthropic PBC v. U.S. Department of Defense. (2026, March 9). Complaint for Declaratory and Injunctive Relief, No. 3:26-cv-01996-RFL. U.S. District Court, Northern District of California.

https://www.courthousenews.com/wp-content/uploads/2026/03/anthropic-supply-chain-risk-lawsuit.pdf

[46] Reuters. (2026f, March 9). Anthropic sues Pentagon over supply chain risk designation. https://www.reuters.com/business/anthropic-sues-pentagon-over-supply-chain-risk-designation-2026-03-09/

[47] OpenAI. (2025, August 5). Introducing gpt-oss. *OpenAI*. https://openai.com/index/introducing-gpt-oss/

[48] Microsoft. (2025, August 5). OpenAI's open-source model: gpt-oss on Azure AI Foundry and Windows AI Foundry. https://azure.microsoft.com/en-us/blog/openais-open%E2%80%91source-model-gpt%E2%80%91oss-on-azure-ai-foundry-and-windows-ai-foundry/

[49] OpenAI. (2025, October 27). Response to the White House Office of Science and Technology Policy request for information on AI regulatory reform. *OpenAI*. https://cdn.openai.com/pdf/21b88bb5-10a3-4566-919d-f9a6b9c3e632/openai-ostp-rfi-oct-27-2025.pdf

[50] OpenAI. (2026, February 28). Our agreement with the Department of War. *OpenAI*. https://openai.com/index/our-agreement-with-the-department-of-war/

[51] Reuters. (2026e, March 3). OpenAI amending deal with Pentagon, CEO Altman says. https://www.reuters.com/business/openai-amending-deal-with-pentagon-ceo-altman-says-2026-03-03/

[52] Associated Press. (2025, March 28). Elon Musk sells X to his own xAI for $33 billion in all-stock deal. *AP News*. https://apnews.com/article/b245f463076ac9b72c41f92160dc77eb

[53] xAI. (2026, February 2). xAI joins SpaceX. https://x.ai/news/xai-joins-spacex

[54] Associated Press. (2026, February 2). Elon Musk merges his rocket company SpaceX with AI startup xAI. *AP News*. https://apnews.com/article/2079f03fa888652b7fe836afe8b670a1

[55] Microsoft. (2024, February 26). Microsoft and Mistral AI announce new partnership to accelerate AI innovation and introduce Mistral Large first on Azure. https://azure.microsoft.com/en-us/blog/microsoft-and-mistral-ai-announce-new-partnership-to-accelerate-ai-innovation-and-introduce-mistral-large-first-on-azure/

[56] Dillet, R. (2024, February 27). Microsoft made a $16 million investment in Mistral AI. *TechCrunch*. https://techcrunch.com/2024/02/27/microsoft-made-a-16-million-investment-in-mistral-ai/

[57] CNBC. (2025, September 9). AI firm Mistral valued at $14 billion as ASML takes major stake. https://www.cnbc.com/2025/09/09/ai-firm-mistral-valued-at-14-billion-as-asml-takes-major-stake.html

[58] Christensen, C. M. (1997). *The innovator's dilemma*. Harvard Business School Press.

[59] Shapiro, C., & Varian, H. R. (1999). *Information rules: A strategic guide to the network economy*. Harvard Business School Press.

[60] Utterback, J. M., & Abernathy, W. J. (1975). A dynamic model of process and product innovation. *Omega, 3*(6), 639–656.

[61] Reuters. (2026c, March 4). Nvidia CEO says firm unlikely to invest $100B in OpenAI as it prepares for IPO. https://www.reuters.com/business/nvidia-will-not-be-able-invest-100-billion-openai-due-ipo-ceo-jensen-says-2026-03-04/

[62] Reuters. (2025c, December 24). Nvidia, joining Big Tech deal spree, to license Groq technology, hire executives. https://www.reuters.com/business/nvidia-buy-ai-chip-startup-groq-about-20-billion-cnbc-re

ports-2025-12-24/

[63] Groq. (2025, December 24). Groq and Nvidia enter non-exclusive inference technology licensing agreement to accelerate AI inference at global scale. https://groq.com/newsroom/groq-and-nvidia-enter -non-exclusive-inference-technology-licensing-agreement-to-accelerate-ai-inference-at-global-scale

[64] Lewis, P., Perez, E., Piktus, A., Petroni, F., Karpukhin, V., Goyal, N., Häusser, M., García, X., Izacard, G., Subramanian, V., Hosseini, A., Dwivedi-Yu, J., Stoyanov, V., Grave, E., Riedel, S., & Kiela, D. (2020). Retrieval-augmented generation for knowledge-intensive NLP tasks. *arXiv:2005.11401*.

[65] Yao, S., Zhao, J., Yu, D., Du, N., Shafran, I., Narasimhan, K., & Cao, Y. (2023). ReAct: Synergizing reasoning and acting in language models. *arXiv:2210.03629*.

[66] Wang, L., Ma, C., Feng, X., Zhang, Z., Yang, H., Zhang, J., Chen, Z., Tang, J., Chen, X., Lin, Y., Zhao, W. X., Wei, Z., & Wen, J.-R. (2024). A survey on large language model based autonomous agents. *arXiv:2308.11432*.

[67] Grogan, J. J. (2025). AgentFacts: Universal KYA standard for verified AI agent metadata and deployment. Universitas AI. *arXiv preprint*. https://arxiv.org/abs/2506.13794

[68] Washington Post. (2026, February 27). How Anthropic and the Pentagon got into a fight over AI weapons. https://www.washingtonpost.com/technology/2026/02/27/anthropic-pentagon-ai-weapons/

[69] Washington Post. (2026, March 4). Anthropic's AI tool Claude central to U.S. campaign in Iran, amid a bitter feud. https://www.washingtonpost.com/technology/2026/03/04/anthropic-ai-iran-campaign/

[70] U.S. Department of War. (2026). Senior officials outline President's proposed FY26 defense budget. https://www.war.gov/News/News-Stories/Article/Article/4227847/senior-officials-outline-presidents-propo sed-fy26-defense-budget/

[71] Reuters. (2026d, January 7). Trump calls for $1.5 trillion military budget in 2027. https://www.reuters.com/world/us/trump-says-us-military-budget-2027-should-be-15-trillion-2026-01-07/

[72] Office of the Director of National Intelligence. (2023). *Annual threat assessment of the U.S. intelligence community*. Washington, D.C. https://www.dni.gov/files/ODNI/documents/assessments/ATA-2023-Unclassified-Report.pdf

[73] Department of Defense. (2024). *Data, analytics, and artificial intelligence adoption strategy*. Washington, D.C. https://media.defense.gov/2023/Nov/02/2003333300/-1/-1/1/DOD_DATA_ANALY TICS_AI_ADOPTION_STRATEGY.PDF

[74] Government Accountability Office. (2024). *Artificial intelligence: Agencies have begun implementation but need to complete key requirements* (GAO-24-106821). Washington, D.C. https://www.gao.gov/products/gao-24-106821

[75] White House. (2025, September 5). *Restoring the United States Department of War* (Executive Order 14347). https://www.whitehouse.gov/presidential-actions/2025/09/restoring-the-united-states-departme nt-of-war/

*Grogan, J. J. (2026). The End of the Foundation Model Era: Open-Weight Models, Sovereign AI, and Inference as Infrastructure. arXiv preprint, Version 1.0. March 9, 2026.*